# $K^0 - \bar{K}^0$ mixing in full lattice QCD

Weonjong Lee [a*] and Markus Klomfass [b]

[a]Department of Physics, Columbia University, New York City, NY 10027, U.S.A.

[b]Veilchenweg 24, 65201 Wiesbaden, Germany.

There are at least two methods to calculate $B_K$ with staggered fermions: one is the two spin trace formalism and the other is the one spin trace formalism. We have performed numerical simulations on a $16^3 \times 40$ lattice in full QCD with $\beta = 5.7$ and a dynamical quark mass 0.01 in lattice units. We try various sources to select only the pseudo-Goldstone bosons and compare the various results.

## 1. INTRODUCTION

The lattice simulation of weak matrix elements has been one of the most significant contribution to Standard Model phenomenology. Especially, the knowledge of $B_K$ which describes $K^0 - \bar{K}^0$ mixing through the $\Delta S = 2$ electro-weak Hamiltonian is important in order to determine precisely the parameters of the Standard Model from experimental data.

There are two methods to transcribe a continuum weak matrix element (*eg* $B_K$) to the lattice with staggered fermions [1,2]: one is the *one spin trace formalism* and the other is the *two spin trace formalism*. By writing the four–fermion operators as a product of operators bilinear in the fermion fields, in the *one spin trace formalism* each external hadron is contracted with both bilinears of the four–fermion operators simultaneously, whereas in the *two spin trace formalism* each external hadron is contracted with only one of the bilinears [1,2]. The problem is that the operators in both formalisms are different on the lattice whereas the operators in both formalisms are identical in the continuum by Fierz transformation. There have been a number of attempts to evaluate $B_K$ on the lattice with staggered fermions using both a Landau gauge operator and a gauge-invariant operator in the *two spin trace formalism* [3].

Here we will explain the *one spin trace formalism* and compare the numerical results between both formalisms. Furthermore we try an improved numerical technique (*cubic wall source*) in order to select only the pseudo-Goldstone boson signal exclusively. The results of the *cubic wall source* are compared with the conventional method (*Kilcup and Sharpe wall source* [4]).

## 2. ONE SPIN TRACE FORMALISM

We follow the notation for the four–fermion operators of Ref. [1,5]. In the continuum, $B_K$ is defined as

$$B_K \equiv \frac{\langle \bar{K}^0 \mid \bar{s}\gamma_\mu(1-\gamma_5)d \; \bar{s}\gamma_\mu(1-\gamma_5)d \mid K^0 \rangle}{\frac{8}{3}\langle \bar{K}^0 \mid \bar{s}\gamma_\mu\gamma_5 d \mid 0\rangle\langle 0 \mid \bar{s}\gamma_\mu\gamma_5 d \mid K^0\rangle} \; . \quad (1)$$

In the *two spin trace formalism*, the four–fermion operator in the numerator in Eq. (1) is transcribed to the lattice [1,5] as a sum of four terms:

$$\begin{aligned} \mathcal{O}^{Latt}_{2TR} &= (V \times P)^{2TR}_{ab;ba} + (V \times P)^{2TR}_{aa;bb} \\ &\quad + (A \times P)^{2TR}_{ab;ba} + (A \times P)^{2TR}_{aa;bb} \end{aligned} \quad (2)$$

where V (or A) represents the vector (or axial) spin structure, P represents the pseudoscalar-like flavor structure and the subscript $ab;ba$ (or $aa;bb$) represents the color indices of the quark fields [1]. The expression in Eq. (2) has the same chiral behavior in the limit of vanishing quark mass as does the continuum operator (this is true even for the analytic parts of each term in Eq. (2) separately) [5,1]. In addition, $\mathcal{O}^{Latt}_{2TR}$ preserves the same leading logarithmic behavior as the continuum $\Delta S = 2$ operator [6,1].

*Research sponsored in part by the U.S. Department of Energy.

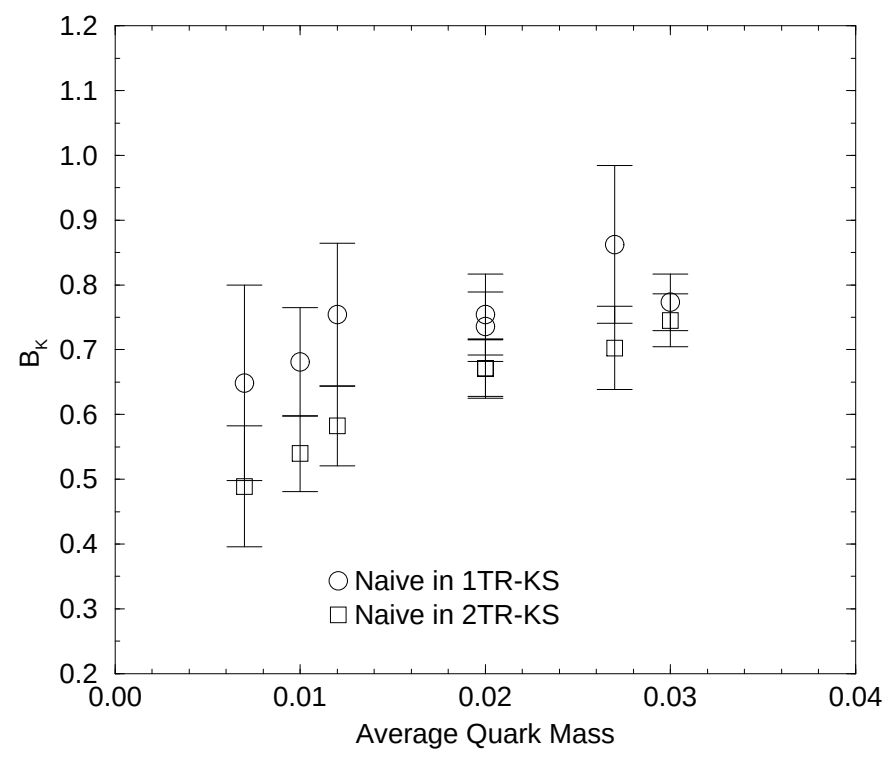

Figure 1. Comparison of naive $B_K$ in both one spin trace and two spin trace formalisms calculated with Kilcup and Sharpe wall source.

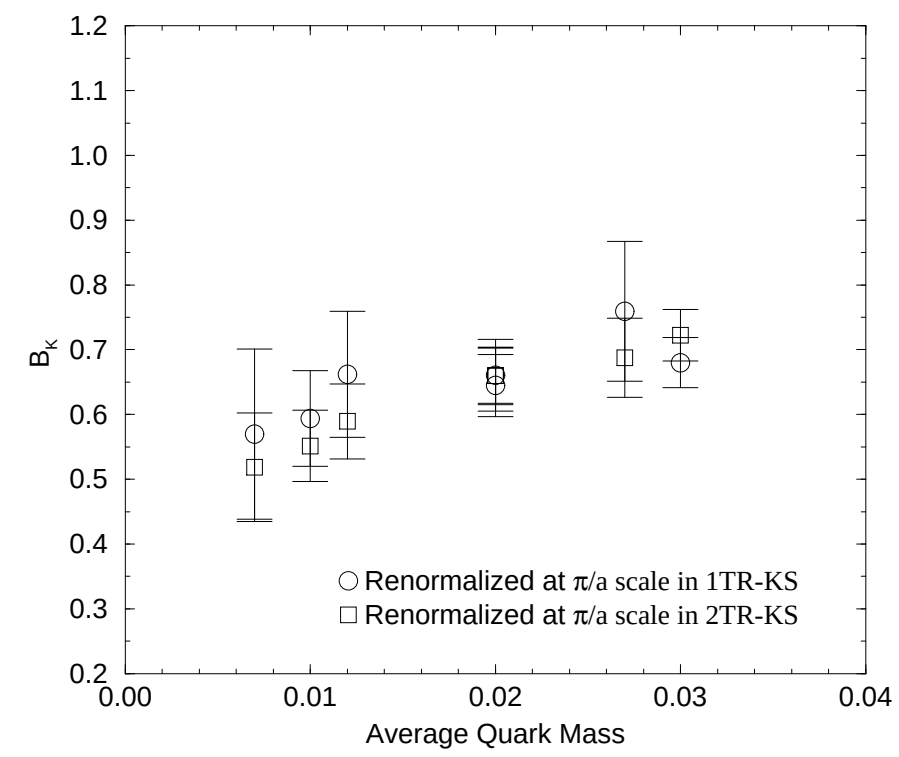

Figure 2. Comparison of tadpole-improved renormalized $B_K$ at $\pi/a$ scale in both one spin trace and two spin trace formalisms calculated with Kilcup and Sharpe wall source.

In the *one spin trace formalism*, the four-fermion operator of the numerator in Eq. (1) is transcribed to the lattice as follows:

$$\mathcal{O}^{Latt}_{1TR} = (V\times P)^{1TR}_{ab;ba} + (V\times P)^{1TR}_{aa;bb} + (A\times P)^{1TR}_{ab;ba} + (A\times P)^{1TR}_{aa;bb} + \mathcal{O}^{1TR}_{\text{chiral partner}} \,. \quad (3)$$

In contrast with the *two spin trace formalism*, the individual terms in Eq. (3) do not possess the same chiral behavior as the continuum $\Delta S = 2$ operator [1]. We must add $\mathcal{O}^{1TR}_{\text{chiral partner}}$ in order to preserve the correct continuum chiral behavior [1]. By imposing the correct chiral behavior on $\mathcal{O}^{Latt}_{1TR}$, we determine the chiral partner operator [1] as follows:

$$\mathcal{O}^{1TR}_{\text{chiral partner}} = (V\times S)^{1TR}_{ab;ba} + (V\times S)^{1TR}_{aa;bb} + (A\times S)^{1TR}_{ab;ba} + (A\times S)^{1TR}_{aa;bb} \,. \quad (4)$$

This forces the resulting operator to respect the continuum chiral behavior. The next question is whether the additional chiral partner operators still have the continuum leading logarithmic behavior. By choosing basis operators which belong to the identity representation with respect to the 90° axial rotation group (a subgroup of the exact $U_A(1)$ symmetry group), we find an eigen-operator (Eq. (3) and Eq. (4)) which possesses the same chiral behavior and the same leading logarithmic behavior as the continuum $\Delta S = 2$ operator [1].

The Fierz transform of $\mathcal{O}^{Latt}_{1TR}$ in Eq. (3) is different from $\mathcal{O}^{Latt}_{2TR}$ in Eq. (2) by the following terms [1]:

$$\mathcal{O}^{Latt}_{1TR} - \mathcal{O}^{Latt}_{2TR} = ((V+A)\times(S-T))^{2TR} \,. \quad (5)$$

The contraction of $((V+A)\times(S-T))^{2TR}$ operator with external pseudo-Goldstone Kaons vanishes as $a \to 0$, since the pseudo-Goldstone Kaon has the pseudoscalar-like flavor matrix and the flavor trace in the contraction is zero. The above argument of vanishing flavor trace is not true for finite lattice spacing $a \neq 0$ since flavor symmetry violation (flavor mixing) is present for $a \neq 0$. This means not only that the difference of the numerical results in both formalisms shows how close we are to the continuum but also that as $a \to 0$ the continuum Fierz transformation property of $\Delta S = 2$ operator is recovered [1].

The numerical simulation of $B_K$ in both formalisms has been calculated using the Columbia 16 Giga-flop parallel processor. Configurations of $16^3 \times 40$ size are updated using R-algorithm with $\beta = 5.7$ ($1/a \cong 2$ GeV) for two dynamical flavors with $m_q a = 0.01$. For the $B_K$ measurement, the configuration is doubled ($16^3 \times 80$) along time direction and the distance between the two wall sources is 36. The quark mass pairs for the Kaon are (0.01,0.01), (0.02,0.02), (0.03,0.03), (0.004,0.01), (0.004,0.02), (0.01,0.03) and (0.004,0.05). Results calculated with a Kilcup and Sharpe source [4] in both formalisms are compared in Figures 1 and 2. The tadpole-improved renormalization for $B_K$ are explained in detail in Ref. [1]. From Figures 1 and 2, we conclude that the tadpole-improved renormalized $B_K$ data in both formalisms are in good agreement with each other.

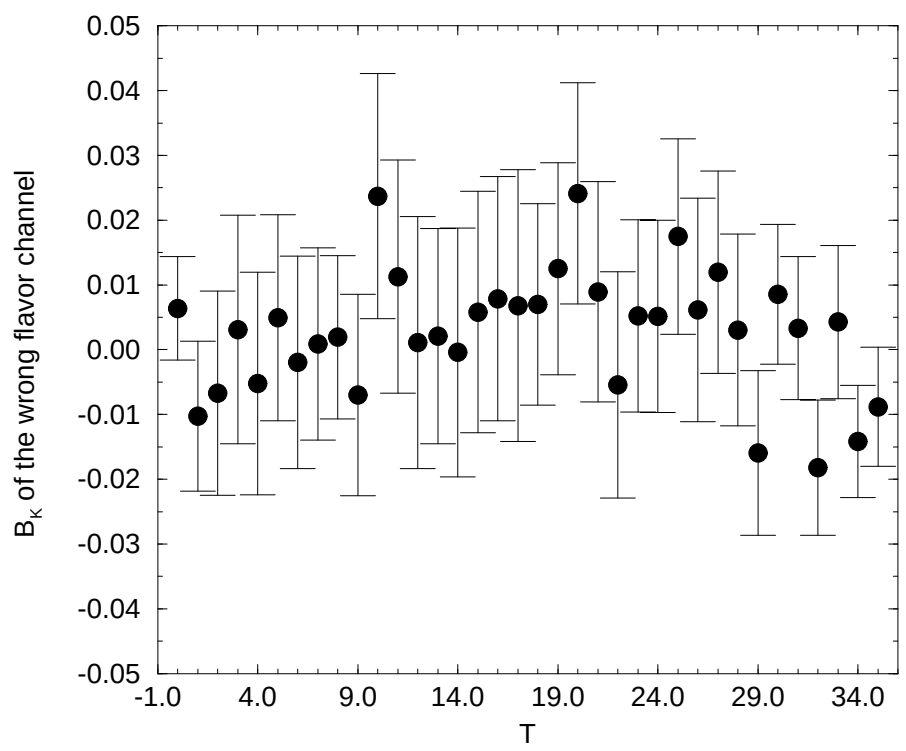

Figure 3. Naive $B_K$ with wrong (scalar-like) flavor structure and valence quark mass 0.02 calculated with *Kilcup and Sharpe wall source* in two spin trace formalism, which are supposed to vanish as $a \to 0$.

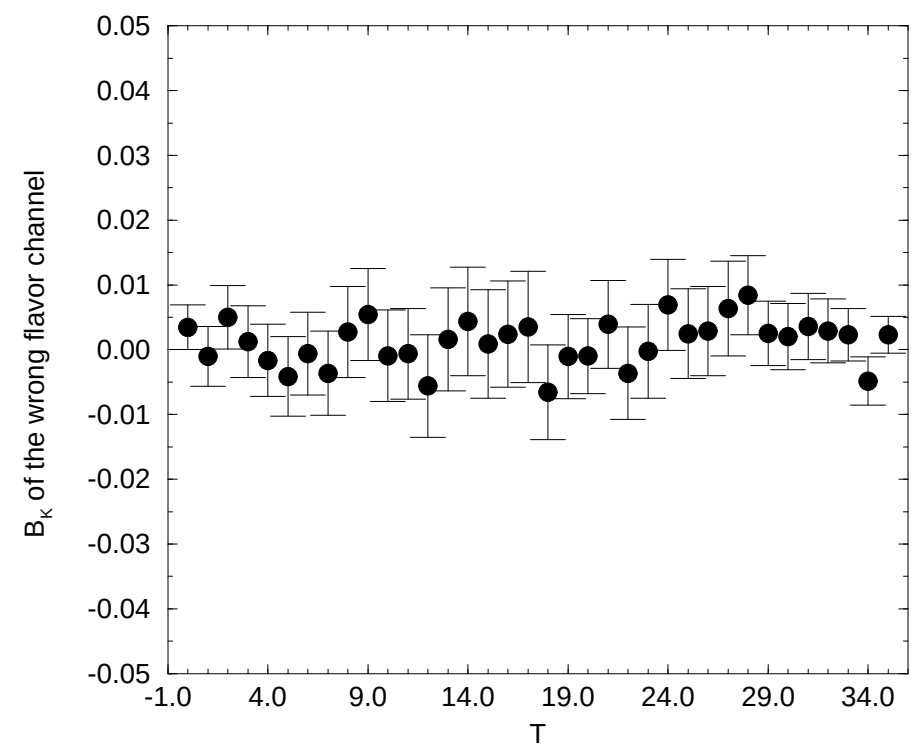

Figure 4. Naive $B_K$ with wrong (scalar-like) flavor structure and valence quark mass 0.02 calculated with *cubic wall source* in two spin trace formalism, which are supposed to vanish as $a \to 0$.

## 3. CUBIC SOURCE METHOD

For hadron spectrum measurements, the sink possesses the same symmetry as a specific hadronic state. In contrast to the hadron spectrum measurements, the weak matrix element measurements require the symmetry properties of the wall source to determine the specific hadronic state since the electro-weak effective Hamiltonian does not select any particular hadronic state by itself. There is a technique (*Kilcup and Sharpe wall source*) which selects only pseudo-Goldstone states out of many pion states [4], but may possibly also produce $\rho$ meson states.

For the *cubic source* method [7], we implement the GRF (Geometric Rest Frame) group on the wall source in such a way that for each quark flavor, 8 propagators are obtained for sources that are non-zero for each corner of the spatial cubes making up the wall. Specific combinations of these 8 propagators are chosen to project on the pseudo-Goldstone mode [7]. We have tried these two kinds of wall sources in our numerical simulation of $B_K$. The manifest improvement given by the cubic source can be seen by examining the wrong flavor channel of $B_K$ $((V+A)\times S)$ which is supposed to vanish in the continuum limit $a = 0$, as in Figures 3 and 4. Our results, obtained on 63 configurations, are naive $B_K^{2TR} = 0.66(6)$, naive $B_K^{1TR} = 0.76(11)$, tadpole-improved $B_K^{2TR}(\frac{\pi}{a}) = 0.66(5)$, and tadpole-improved $B_K^{1TR}(\frac{\pi}{a}) = 0.67(10)$.

Kind help from Prof. N. Christ, Prof. R. Mawhinney, D. Zhu, S. Chandrasekharan and D. Chen is acknowledged with gratitude.